\documentstyle[sprocl]{article}

\def\be{\begin{equation}}
\def\ee{\end{equation}}
\def\bea{\begin{eqnarray}}
\def\eea{\end{eqnarray}}

\newcommand{\cA}{{\cal A}}

\newcommand{\bi}{\bigskip}
\newcommand{\no}{\noindent}
\newcommand{\hk}{\hspace{0.1cm}}
\newcommand{\hs}{\hspace{0.5cm}}

\newcommand{\rk}{\right)}
\newcommand{\lk}{\left(}

\newcommand{\sli}{\sum\limits}

\newcommand{\il}{\int\limits}

\begin{document}

\title{Magnetic Monopoles, Vortices and the Topology of Gauge
Fields\footnote{Supported in part by Deutsche Forschungsgemeinschaft under
contract no.~DFG-Re~856/4-1}\footnote{Invited talk given at the International
Workshop ``Understanding Deconfinement in QCD'', Trento, 1.-13. March, 1999}}

\author{H. Reinhardt}

\address{Institute for Theoretical Physics\\
 T\"ubingen University}

\maketitle\abstracts{ Lattice calculations performed in Abelian gauges give strong evidence that
confinement is realized as a dual Meissner effect, implying that the Yang-Mills
vacuum consists of a condensate of  magnetic monopoles. Alternative lattice
calculations performed in the maximum center gauge give strong support that
center vortex configurations are the relevant infrared degrees of freedom
responsible for confinement and that the magnetic monopoles are mostly sitting
on vortices. In this talk I study the continuum Yang-Mills-theory in Abelian and
center gauges. In Polyakov gauge the Pontryagin index of the gauge field is
expressed by the magnetic monopole charges. The continuum analogues of center
vortices and the continuum version of the maximum center gauge are presented. It
is shown that the Pontryagin index of center vortices is given by their 
self-intersection number, which vanishes unless magnetic monopole currents 
are flowing
on the vortices.}

\section{Introduction}
There are two fundamental essentially non-perturbative features of QCD:
confinement and spontaneous breaking of chiral symmetry. The latter can be
more or less understood in terms of instantons. 
By the Atiah-Singer theorem\cite{1}
instanton fields having a
non-trivial Pontryagin index give 
rise to zero modes
of the quarks and after averaging over all gauge fields favour a non-zero quark
level density $\rho (\lambda) \neq 0$ at zero virtuality $\lambda = 0$.
By the Banks-Casher theorem\cite{2} 
$\langle\bar{q} q \rangle = \pi \rho (0)$ this implies a
non-zero quark condensate, which is the order parameter of spontaneous breaking
of chiral symmetry. However, one should stress that this explanation of
spontaneous breaking of chiral symmetry does not really rely on instantons, i.e. 
on finite action solutions of the classical field equation but only on
topologically non-trivial field configurations.
\bi

\no 
On the other hand the confinement mechanism is much less understood. However,
lattice calculations performed over the last couple of years\cite{3,4}, 
have accumulated
evidence that confinement is realized either as dual Meissner effect\cite{5}
 implying a
condensation of magnetic monopoles in the QCD vacuum or by a condensation of
magnetic vortices\cite{6}. 
In fact, recent lattice calculations also indicate that the
magnetic monopoles are related to the vortices\cite{4}.
\bi

\no
>From the conceptional point of view it is not very appealing that spontaneous
breaking of chiral symmetry and confinement are attributed to different types of
field configurations. This is because lattice calculations indicate that
 the deconfinement
phase transition is accompanied by the restauration of chiral symmetry. Moreover,
the spontaneous breaking of chiral symmetry is determined by topologically
non-trivial field configurations. Since magnetic monopoles are long-range 
fields
they should be relevant for the global topological properties of gauge fields.
One can therefore expect an intimate relation between magnetic monopoles and the
topology of gauge fields. In this lecture I will show that in Abelian gauges the
non-trivial topology of gauge fields is generated by magnetic monopoles. In
these
gauges instantons give rise to magnetic monopoles but the latter do exist
without instantons. Furthermore, magnetic vortices can be
topologically non-trivial only if they host magnetic monopoles. In this sense,
magnetic monopoles have to be considered as the fundamental topological objects
of gauge fields, at least in Abelian gauges.
\bi

\no
\section{Emergence of Magnetic Monopoles in Abelian Gauges}
\bi

\no
Magnetic monopoles arise in the so-called Abelian gauges\cite{8} 
which fix the coset
$G/H$ of the gauge group $G$, but leave Abelian gauge invariance with
respect to the Cartan subgroup $H$. Recent lattice calculations show
that the dual Meissner effect is equally well realized in all Abelian gauges
studied\cite{7}, while the Abelian and monopole dominance is more pronounced in
the so-called maximum Abelian gauge\cite{3}. Here, for simplicity, we will use
the Polyakov gauge, defined by a diagonalization of the  Polyakov loop
\be
\label{1}
\Omega (\vec{x})  =  P \exp \lk - \int \limits_0^T d x_0\, A_0 \rk 
= V^\dagger \omega V \to \omega \hk ,
\ee
where $\omega \in H$ is the diagonal part of the 
Polyakov loop and the matrix $V \in G/H$ is
obviously defined only up to an Abelian gauge transformation $V \to gV$\,, 
$g \in H$. This gauge is equivalent to the condition
$A^{ch}_{0} = 0$\,, $\partial_0 A^n_0 = 0$\,,
where $A^n_0$ and $A^{ch}_0$ denote the diagonal (neutral with respect to $H$)
and off-diagonal (charged) parts of the gauge field, respectively. For
simplicity, we will assume $G = SU (2)$ below.
\bi

\no
When the Polyakov loop (\ref{1}) becomes a center element of the gauge group at
some isolated point $\bar{x_i}$ in 3-space
\be
\label{4}
\Omega (\bar{x}_i) = (- 1)^{n_i} \hk ,
\ee
there is a topological obstruction in the diagonalization and the induced gauge
field
\be
\label{5}
\cA = V \partial V^\dagger \hk ,
\ee
arising from the gauge transformation $V \in G/H$ which makes the Polyakov loop
diagonal, develops a magnetic monopole\cite{8}.
An important property of the magnetic monopoles arising in the Abelian gauges is
that their magnetic charge
\be
\label{6}
m [V] = \frac{1}{4 \pi} \int\limits_{\bar{S}_2} d \vec{\Sigma}\,
\vec{B}^3
\ee
is topologically quantized and given by 
$m [V] \in \Pi_2 (SU(2) / U (1))$, i.e.~by the winding number of the 
mapping $V (\vec{x}) \in SU (2) / U (1)$\cite{8,9}.
In the above equation $\bar{S}_2$ is an infinitesimal 2-sphere around
the monopole position with the piercing point of the Dirac string left out.
\bi

\no
\section{Magnetic Monopoles as Sources of Non-Trivial Topology}
\bi

\no
Since the magnetic monopoles are long-range fields, we expect that they are
relevant for the topological properties of gauge fields. As is well known, the
gauge fields $A_\mu (x)$ are topologically classified by the Pontryagin index
\be
\label{7}
\nu = - \frac{1}{16 \pi^2} \int d^4 x\,\mbox{Tr}\, F_{\mu \nu} 
\tilde{F}_{\mu \nu} \hk .
\ee
In the gauge $\partial_0 A_0 = 0$ (which is satisfied by the Polyakov gauge) and
for temporally periodic spatial components of the gauge field $\vec{A} (\vec{x},
T) = \vec{A} (\vec{x}, 0)$, the Pontryagin index $\nu [A]$ (\ref{7}) equals the
winding number of the Polyakov loop $\Omega (\bar{x})$ (\ref{1}):
\bea
\label{8}
n [\Omega] = \frac{1}{24 \pi^2} \int d^3 x\, \epsilon_{ijk} \,\mbox{Tr}\, L_i L_j
L_k \hk , \hs L_i = \Omega \partial_i \Omega^\dagger \hk .
\eea
For this winding number to be well defined, $\Omega (\vec{x})$ has to approach at
spatial infinity, $\vert \vec{x} \vert \to \infty$, a value independent of the
direction $\hat{x}$, so that the 3-space $R^3$ can be topologically compactified
to the 3-sphere $S_3$. Without loss of generality we can assume that for 
$\vert \vec{x} \vert \to \infty$\,, $\Omega
(\vec{x})$ approaches a center element 
$\Omega (\vec{x}) \longrightarrow  
(-1)^{n_0}$
with $n_0$ integer. In 
ref.~\cite{9} the following exact relation for the winding
number $n [\Omega]$ has been derived
\be
\label{10}
n [\Omega] = \sli_i \ell_i m_i \hk .
\ee
Here the summation runs over all magnetic monopoles, $m_i$ being the monopole
charges, and the integer
\be
\label{11}
\ell_i = n_i - n_0
\ee
is defined by the center element $\Omega (\bar{x}_i) = (-1)^{n_i}$ which the
Polyakov loop aquires at the monopole position $\bar{x}_i$ (cf.~eq.~(\ref{8})),
and by the boundary condition to $\Omega (\vec{x})$ specified above. 
The quantity $\ell_i$ represents the
invariant
length of the Dirac string in group space, i.e.~the distance in group space
between the center elements taken by the Polyakov loop at the monopoles 
connected by the Dirac string.
\bi

\no
Eq.~(\ref{10}) shows that in the Polyakov gauge the topology of gauge fields is
exclusively determined by the magnetic monopoles. Therefore, magnetic
monopoles should be also sufficient to trigger spontaneous breaking of chiral
symmetry, which is usually attributed to instantons. Instantons in Abelian
gauges give rise to magnetic monopoles, but monopoles can exist in the absence
of instantons.
The above considerations show that in the Abelian gauges both confinement and
spontaneous breaking of chiral symmetry can be generated by the same field
configurations: magnetic monopoles.
\bi

\no
Although the above considerations have been explicitly carried out in Polyakov
gauge, I strongly believe that also in other Abelian gauges the Pontryagin index
is given by magnetic charges. In fact, a direct relation between the
topological charge and the magnetic monopole charges has been also seen in
lattice calculations in the maximum Abelian gauge\cite{10,11}. These 
lattice calculations show that the Pontryagin index vanishes when the monopole
part of the Abelian gauge field is removed\cite{10}. 
 Finally let
me mention that a relation simular to eq.~(\ref{10}) has been subsequently 
derived in 
refs.~\cite{12,13}. The precise connection between refs.~\cite{12,13}
 and ref.~\cite{9} has been established in
ref.~\cite{14}. 
\bi

\no
\section{Center vortices}
\bi

\no
The center $Z (N)$ of the gauge group is known to play a crucial role for the
confinement of charges in the fundamental representation\cite{6}. 
In the so-called
maximum center gauge\cite{4} 
the gauge freedom is used to bring the link variables
$U_\mu (x)$ as close as possible to a center element of the gauge group. After
fixing the coset $SU (N) / Z (N)$\,, leaving the center symmetry $Z (N)$ 
untouched,
center projection of the links implies for $SU (2)$ replacing the link variables
$U_\mu (x)$ by their sign. Thereby vortices arise as strings (in $D = 3$) and
sheets (in $D = 4$) of plaquettes
 $(-1)$, which are closed by the Bianchi identity.
The remarkable lattice result is that center projection reproduces the full
string tension\cite{4} 
which has been referred to as center dominance and implies vortex
dominance. In fact, removing the field configurations which after center
projection result in center vortices removes completely the string tension. In
this sense center vortices have been interpreted as the confiners of the theory.
Furthermore, the center vortex picture gives also a natural explanation of the
deconfinement phase transition\cite{15}. 
It has been also shown, that  removal of the vortex configurations destroys the
spontaneous breaking of chiral symmetry\cite{16}.
\bi

\no
Center dominance without center gauge fixing is trivial\cite{17}. 
Therefore, the maximum
center gauge  fixing seems to accumulate the dominant infrared physics on the
vortices. To illustrate the effect of the maximum center gauge fixing 
let us consider
the following Abelian vortex field (in cylindric coordinates $\rho, \varphi, z$)
\be
\label{13}
\vec{a} = \frac{1}{\rho} \,\vec{e}_\varphi\, T_3 \hk ,
\ee
which represents a singular magnetic flux line on the $z$-axis. (Ignoring the
group generator $T_3$ this field represents the gauge potential of a thin
solenoid.) Putting this field configuration on the lattice the maximum center
gauge concentrates the gauge potential on a sheet of plaquettes $(-1)$ bounded by
the vortex at $\rho = 0$ with all other links equal to 1. This configuration is
not changed by center projection, while center projection before maximum center
gauge fixing would remove the vortex totally. This illustrates the important
role of the maximum center gauge fixing before center projection. In the
continuum limit the maximal center gauge fixed field configuration corresponding
to (\ref{13}) becomes 
\be
\label{14}
{\cA} = 2 \pi \delta (y) \Theta (x) \hk ,
\ee
i.e.~the gauge potential is concentrated after maximal center gauge fixing
on a singular sheet (given here 
by the right half of the $x$-$z$-plane), which is bounded
by the vortex. We will refer to this center vortex arising after maximum center
gauge  fixing as ideal center vortex.
\bi

\no
In $D = 4$ the ideal center vortices arising after 
center projection are given by closed magnetic flux sheets $S = \partial
\Sigma$ with all links in the enclosed 3-dimensional volume $\Sigma$ being $(-1)$. In
the continuum these ideal vortices are given by
\be
\label{15}
{\cA}_\mu (k, \Sigma, x) = E (k) \il_\Sigma d^3 \tilde{\sigma}_\mu\, \delta (x
- \bar{x} (\sigma)) \hk ,
\ee
where $\bar{x}_\mu (\sigma)$ denotes a parametrization of the volume $\Sigma$
enclosed by the vortex sheet $S = \partial \Sigma$. Furthermore, $d^3
\tilde{\sigma}_\mu = \frac{1}{3 !} \epsilon_{\mu \alpha \beta \gamma} d^3
\sigma_{\alpha \beta \gamma}$\,, $d^3 \sigma_{\alpha \beta \gamma}$ being the
3-dimensional volume element and $E (k) = E_a (k) T_a$ denotes a vector of the
root lattice of $SU (N) / Z (N)$, which lives in the Cartan subalgebra and whose
exponent gives rise to a center element 
\be
\label{16}
\exp (- E (k)) = Z (k) \in Z (N) \hk , \hs k = 0, 1, 2, \dots, N - 1 \hk .
\ee
The ideal vortex field ${\cA} (k, \Sigma, x)$ indeed contributes the center
element $Z (k)$ to each Wilson loop $C$ non-trivially linked to the vortex $S =
\partial \Sigma$
\be
\label{17}
W [{\cA}] (C)  =  \exp \left[ - \oint\limits_C d x_\mu\, {\cA}_\mu ( k, \Sigma,
x) \right]
 =  \exp \left[ - E (k) I (C, \Sigma) \right]
= Z (k)^{I (C, \Sigma)} \hk ,
\ee
where
\be
\label{18}
I (C, \Sigma) = \oint\limits_C d x_\mu \il_\Sigma d^3 \tilde{\sigma}_\mu\,
\delta^4 (x - \bar{x} (\sigma))
\ee
is the intersection number between $C$ and $\Sigma$.
Performing the Abelian gauge transformation 
$\cA_\mu \to \cA^{V (k, \Sigma)}_\mu$\,, $V (k, \Sigma) = \exp \lk - E (k) 
\Omega(\Sigma, x) \rk$\,, where 
$\Omega (\Sigma, x)$
is the solid angle in $D = 4$\,, the ideal vortex ${\cA}(k, \Sigma, x)$ is
converted into the thin vortex
\be
\label{20}
a_\mu (k, \partial \Sigma, x)  =  \cA_\mu + E (k) \partial_\mu \Omega = E (k)
\il_{\partial \Sigma} d^2 \tilde{\sigma}_{\mu k}\,\partial_k D (x - \bar{x}
(\sigma))
 \hk ,
\ee
where $- \partial_\mu \partial_\mu D (x - x') = \delta^4 (x - x')$. 
Eq.~(\ref{20})
is the $D = 4$ generalization of eq.~(\ref{13}) for arbitrary vortex shapes
$S$. Unlike the ideal center vortex field $\cA (k,  \Sigma, x)$ the thin
vortex field $a_\mu (k, \Sigma, x)$ does not depend on the precise shape of the
hypersurface $\Sigma$ but depends only on its boundary $S = \partial \Sigma$\,,
i.e.~on the flux sheet of the vortex. Since $a_\mu (k, \Sigma, x)$ is gauge
equivalent to $\cA (k,  \Sigma, x)$ it yields the same Wilson loop as it is
immediately seen by noticing that
\be
\label{21}
\oint\limits_C d x_\mu\, a_\mu (k, \partial \Sigma, x) = E (k) L (C, \partial
\Sigma) \hk ,
\ee
where
$L (C, \partial \Sigma) $
is the linking number between $C$ and $S = \partial \Sigma$, which equals the
intersection number $I (C, \Sigma)$.
\bi

\no
A careful analysis\cite{18} shows that the continuum analogue of the maximum center gauge
is given by the condition
\be
\label{23}
- \il \,\mbox{tr}\, \lk A^g_\mu + a_\mu (k, \partial \Sigma) \rk^2 \to \mbox{min} \hk ,
\ee
where the minimalization is performed with respect to all coset 
gauge transformations
$g \in SU (N) / Z (N)$ 
and with respect to all vortex fields $a_\mu (k, \partial 
\Sigma, x)$. Note that
the gauge condition depends only on the thin vortex $a (k, \partial \Sigma)$. 
For fixed
$a_\mu (k, \partial 
\Sigma, x)$ minimalization with respect to all gauge transformations
$g$ 
leads to the background gauge condition
\be
\label{24}
\left[ \partial_\mu - a_\mu (k, \partial \Sigma, x ), A_\mu (x) \right] = 0 \hs
\ee
with the thin vortex figuring as background field. To arrive at eq.~(\ref{24})
 we have
used that $\partial_\mu a_\mu (k, \partial \Sigma, x) = 0$.
\bi

\no
The continuum version (\ref{23} )
shows that the maximum center gauge condition brings a
given gauge potential as close as possible to a center vortex field, which we
can either represent as a thin vortex $a_\mu (k, \partial 
\Sigma, x)$ or as an ideal
vortex $\cA (k, \Sigma, x)$. The latter is the direct analogue of the center
vortices arising on the lattice after maximum center gauge fixing and center
projection. There is, however, an important difference between the ideal center
vortices in the continuum and on the lattice. Contrary to $\cA (k, \Sigma,
x)$ the lattice center vortices defined after center projection by 3-dimensional
volumes $\Sigma$ of links $(-1)$ do not distinguish between the direction of the
magnetic flux on $S = \partial \Sigma$, i.e.~the lattice vortex sheets  $S =
\partial \Sigma$ are not oriented. 
As a consequence the center projection on the lattice
removes topological properties of the vortices related to their 
orientations, which
enter the Pontryagin index.
\bi

\no
In the continuum theory the field strength of (thin or ideal) center vortices is
given by
\be
\label{25}
F_{\mu \nu} [\cA] = E (k) \il_{\partial \Sigma} d^2 \tilde{\sigma}_{\mu \nu}\,
\delta^4 (x - \bar{x} (\sigma)) \hk , \hs d^2 \tilde{\sigma}_{\mu \nu} = \frac{1}{2}
\epsilon_{\mu \nu \kappa \lambda} d \sigma_{\kappa \lambda} \hk .
\ee
From this expression one finds for the Pontryagin index of a center vortex
\be
\label{26}
\nu [\cA] = \nu [a] = \frac{1}{4} I (S, S)\hk ,
\ee
where $I (S, S)$ is the self intersection number of the vortex sheet $S =
\partial \Sigma$. A more detailed analysis shows that the Pontryagin index is
indeed integer valued and vanishes unless there are magnetic monopole currents
flowing on the vortex sheet. Thus even in the vortex picture the non-trivial
topology is generated by magnetic monopole loops in agreement with the findings
in the Polyakov gauge given above. 
\bi

\no
{\bf Acknowledgment:}\\
I thank M.~Engelhardt, K.~Langfeld, M.~Quandt, A.~Sch\"afke and O.~Tennert 
for collaboration on various aspects of this work.

\end{document}